# DOWNSIZING FROM THE POINT OF VIEW OF MERGING MODEL
## (preliminary discussion)

V.M. Kontorovich[1,2]

[1] *Institute of Radio Astronomy NASU*
*4 Chervonopraporna Str., Kharkov, 61002, Ukraine*,
[2] *Kharkov National V.N. Karazin University*
*4, Svobody square, Kharkov 61022, Ukraine*,

E-mail:vkont@rian.kharkov.ua

In four-particle processes of scattering with transfer of mass, unlike mergers in which mass can only increase, mass of the most massive galaxies may be reduced. Elementary model describing such process is considered. In this way, it is supposed to explain observed phenomenon of "dawnsizing" when increasing of characteristic mass the heaviest galaxies over cosmological time replaces by its reduction.

## 1. Introduction

Quite unusual in terms of the paradigm of mergers, but has long discussed the fact that the maximal galaxy masses (Shechter parameter ), which grow up with decreasing of the red shift at large distances, begin to decrease as you get closer to the present time (see. Figure 1), seems be in contrary to the model of mergers. We will show that it is not.

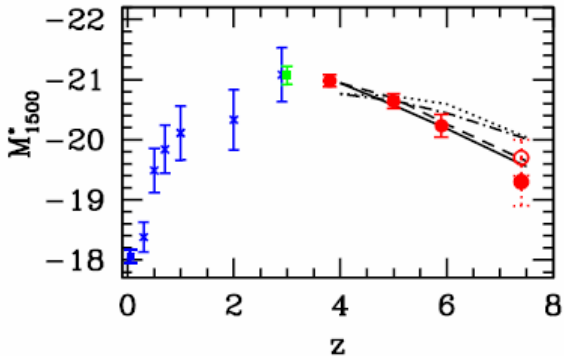

FIG. 13.— Evolution of the characteristic luminosity ($M^*$) of the UV LF as a function of redshift. Determinations are from the present work (red circles) at $z \sim 4-6$, Steidel et al. (1999) at $z \sim 3$ (green square), Arnouts et al. (2005) (blue crosses) at $0.1 \lesssim z \lesssim 3$, and Wyder et al. (2005) at $z \lesssim 0.1$ (blue square).

**Figure 1**. Observational data of the Hubble ultra deep field [1] relating to the Shechter parameter M* (in our consideration it is the relevant characteristical maximal masses).

In the model of galaxy mergers, built on the basis of Smoluchowski kinetic equation (KE), are taken into account only processes of the (paired) mergers, that is the processes involving three "particles" (App.A, Figure 2). The resulting solutions (App.B) allow to find the slope of the mass function $f(M,t)$ in a wide range of redshifts [2,3], satisfactorily explain the observational data of the Hubble ultra deep field [1] (the evolution of the slopes until limiting redshifts). However, arising at this "explosive" evolution leads to an unlimited growth of the maximal mass as we approach the time of the "explosion" [2]. "Explosive" singularity in the solution manifests itself as unlimited growth of the MF as we approach the time of the "explosion" $t=t_{cr}$. Asymptotics of solution $K(M, t)$ for the modifide MF (mMF, App B) $F = M^u f$ near the singularity (outside the physical area of the power behavior) has the form (for $u$=2)

$$K(M,t) \to \frac{\beta}{\sqrt{\frac{1}{M} - \frac{1}{M_{\max}(t)}}},$$

$$M_{\max}(t) = \frac{1}{c(t_{cr} - t)} \qquad (1)$$

That artefact, associated with the use of instant $\delta$-shaped source in the Smoluchowski KE, possible to avoid by the obvious physical regularization [2,3], the meaning of which is in taking into account a finite rise time of gravitational instability, which leads to separation galaxies from the general expansion of the universe. Mathematically, this was taken into account by blur of $\delta$-function on the right side of KE and its replacement by a $\Pi$-shaped stepping stone with a finite small width $\Delta$. The values of MF remain finite in the region of maximal masses too. But still, the very maximal mass in regularized solutions also increases indefinitely when approaching the moment of the explosion [2-4].

As in other similar tasks, taking into account three-particle processes (in our case - mergers of galaxies), leading to the explosive evolution, the finite results occur when four-particle processes come into play in the vicinity of singularity and, in our case, describe the scattering with transfer of mass. In this case, unlike mergers (Figure 2), in which the mass can only increase, a significant role is played the scattering processes in which the mass of the most massive galaxies may be decrease (Figure 3, 4). Below, we consider a simple model scheme describing the disaggregation. In this way, it is supposed to explain the observed "downsizing" phenomenon (Figure 1), where increasing the heaviest characteristic mass over time replaced of its decreasing.

## 2. KE WITH SCATTERING

We remain, as in [2], restrict ourselves by a differential approach, which describes the transfer of a small mass. But now the kinetic equation from the linear is converted into a non-linear (quasi-linear), the most simple form of which is to occur in the KE the nonlinear term $-\tilde{\gamma} F \partial F / \partial M$, where the coefficient $\tilde{\gamma}$ indicates the probability of the "inelastic" scattering process.

We restrict ourselves initially by a merging probability proportional to the square of the mass $M^2$. In this case of the simplest model is natural to choose the same dependence on the mass for the probability of scattering too: $\tilde{\gamma} = \gamma M^2$. To do this, there are physical reasons which we do not discuss here. By introducing variable $z = M^{-1}$ we rewrite quasilinear term $-\gamma M^2 F \partial F / \partial M$ in form $\gamma F \partial F / \partial z$. Although the source in KE quite substantial, mentioned asymptotic expression (1) satisfies the homogeneous kinetic equation, which we confine ourselves.

Our problem [1] thus reduces to the solution of the differential equation

$$\frac{\partial F}{\partial x} + g(F) \frac{\partial F}{\partial z} = 0, \qquad (2)$$

where $g(F) = C\Pi + \gamma F$ is linear on mass function $F$. The equation (2) is a generalized Hopf equation and very well studied. The solution of Cauchy problem of that KE for the mass function $F(M,t)$ with the quasi-linear term having a coefficient $g(F)$ [5], reduces to cubic equation[2] ($x$- time $t$-$t_0$ where $t_0$ – the moment of linking with an explosive solution, playing the role of the initial conditions of the Cauchy problem for the equation (3), $z = M^{-1}$ where $M$ – the mass of galaxy):

$$\gamma(t-t_0)F^3 - \left\{[\frac{1}{M} - \frac{1}{M_0}] - C\Pi \cdot (t-t_0)\right\} F^2 + \beta^2 = 0 \qquad (3)$$

Here $\gamma$ – a nonlinearity parameter, $C\Pi$ – parameter entered from solving a linear KE, namely, $C$ – factor in the probability of mergers of galaxies $CM^2$, $\Pi = \int_0^\infty dM_2 M_2 f(M_2)$ – total mass of low-mass galaxies, $M_0 \equiv M_{max}(t_0)$ – the maximal mass of the galaxies in the linear theory [2] at a time $t_0$, $\beta$ – parameter of the asymptotics of solution of linear problem (1), which is used as an initial condition for solving KE with the nonlinear term.

At $t = t_0$ MF $F(M,t)$, as follows from (3) satisfies the initial condition (1)

$$F^2(M,t_0) = \frac{\beta^2}{[\frac{1}{M} - \frac{1}{M_0}]}, \qquad (4)$$

(corresponding to the asymptotics of our explosive solution of the linear KE), and the moment $t_0$ was selected close to $t_{cr}$, in order to be able use a simple analytical form of the asymptotics of (1)). With $M$, close to $M_0$ – the maximal mass of the linear solution at the moment $t_0$, – it is a large value. (Excluding non-linearity, it tends to infinity at a time approaching то the moment of explosion $t \to t_{cr}$)

## 3. SOLUTION OF KE WITH SCATTERING

We are interested in a real solution of the cubic equation (3) for $F(M,t)$ on large time $t \gg t_0$ as a function of $M$, in particular, the behavior of the new nonlinear "maximal mass", which is yet to be determined, and its dependence on time

We restrict ourselves to demonstrating the asymptotic solution of the cubic equation (3) for the mMF at time $t \gg t_0$ and masses $M \ll M_0$. It is found neglecting the free term in (3). For $\gamma > 0$ there is a unique solution corresponding negative curly brackets

$$F \approx \frac{\frac{1}{M} - \frac{1}{M_0} - C\Pi \cdot (t-t_0)}{\gamma \cdot (t-t_0)} \to \frac{\frac{1}{M} - C\Pi \cdot t}{\gamma \cdot t}. \qquad (5)$$

And the mass is bounded from above by (vanishing curly brackets (3))

$$M < M_{max}(t) = \frac{1}{C\Pi \cdot t}. \qquad (6)$$

It can be seen that the maximal mass decreases with time, what is the required dawnsizing phenomenon. In the resulting solution all the quantities are finite[3]. From the explosive evolution remained only a local on the masses increasing of MF solution near the former peculiarity. It should be noted that this fact may serve as evidence of the explosive stage of evolution.

## 4. DISCUSSION

Thus, a complete solution is a decreasing power function, such as Schechter function, which, however, before the recession in large masses begins to increase at times close to the time of the "explosion" [2,3]. The intervention of processes downsizing (scattering with decreasing of maximal mass) leads to a decrease as both MF with increasing of mass, which passes through a local maximum, and descending of the maximal mass with time. This is consistent with the observed effect of downsizing.

---

[1] We use the notation of the reference book by Zaitsev and Polyanin [5], item 12.4.2.1, point 2 (p. 271)
[2] The latter is easily verified by its direct differentiation on time and on mass.
[3] With the exception of infinity introduced by the initial condition. Starting from the regularized solution (if $t < t_0$) we would get the final values. But this leads to a more cumbersome calculations.

From dependence $M^2$ in the probability of mergers easily go to any power dependence $M^u$ by replacement $z = M^{1-u}/(u-1)$, ie by replacement $M^{-1} \to M^{1-u}/(u-1)$ in the resulting solution (cf [2]).

We were deciding an extremely simplified model problem. But even in such a setting appears the phenomenon downsizing. In reality, the process of downsizing should be described by integral kinetic equation, when as a result of scattering galaxies appear the galaxies of comparable masses.

Author is grateful to Boris Komberg for debates, which led the author to discussion the problem, to Alexander Kats for participation in previous joint works on the theme, as well as Dr. R. J. Bouwens and his co-authors for permission to reproduce the Figure from the paper [1].

## 4. APPENDIX A

Following are the schematic drawings for explaining considered above the merger and scattering.processes

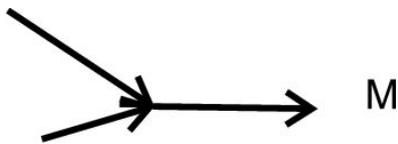

**Figure 2**. The process of mergers by the triple processes with mass increasing, leading to Smoluchowski KE. At low mass transfer KE becomes differential [2,3]. The processes shown in Figures 3 and 4, under conditions of low mass transfer leads to the considered quasilinear KE (2) describing the downsizing. Through M in all figures denotes the mass of the most massive galaxy.

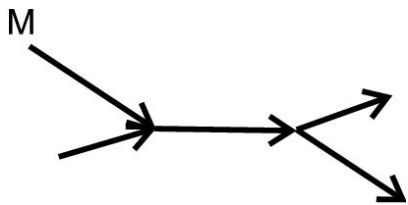

**Figure 3**. The process of merging with the appearance of an unstable intermediate galaxy, which immediately disintegrates. (Effective scattering due to the triple процесс in the second order). And the highest mass of galaxies can be reduced, which leads to the downscaling.

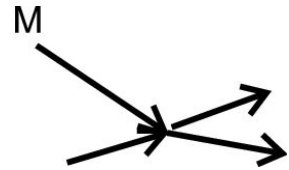

**Figure 4**. The direct scattering by quaternary processes with downsizing of galaxies. In fact, the equation we use corresponds to mass loss during its transmission. In the absence of losses occurs nonlinear equation somewhat different type. Let us note, that accounting of losses in merger process are not essential and does not lead to qualitative effects.

## 5. APPENDIX B

Consider solutions of the Smoluchowski KE in the differential form supposing that the main contribution is due to mergers of the low-mass galaxies with the massive ones with the corresponding merging probability, $U(M_1, M_2) \simeq 0.5CM_1^u$ for $M_2 \ll M_1$.

$$\frac{\partial}{\partial t}f(M,t) + C\Pi\frac{\partial}{\partial M}\left[M^u f(M,t)\right] = \phi(M,t). \quad (B1)$$

Rewriting Eq. (1) for mMF $F(M,t) = M^u f(M,t)$, as

$$\frac{\partial}{\partial t}F(M,t) + C\Pi M^u \frac{\partial}{\partial M}F(M,t) = \Phi(M,t), \quad (B2)$$

where the modified source is $\Phi(M,t) = M^u \phi(M,t)$, we restrict themselves by the localised sourse, that give us the possibility to find solution explicitly [2-3]. The solution for MF has the power-low part, which is in good agremant with observational data. But near the maximal mass the MF has non physical singuliarities. Regularization [2,3] leads to the Shechter type MF, that has no singuliarities but the maximal mass tends to infinity when the moment of time goes to the explosion time $t_{cr}$.

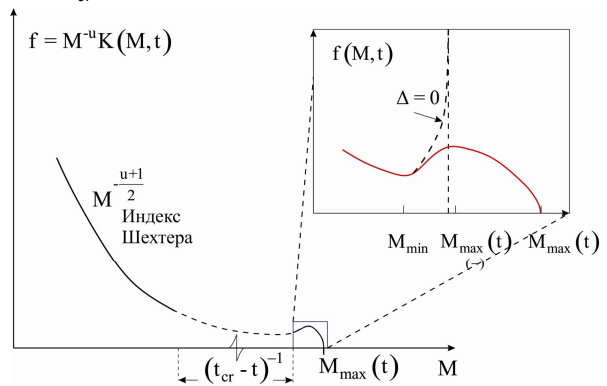

**Figure 5**. The MF established as a result of only mergers (Fig.2) with small mass increments. The dashed line shows the MF singularity in the case of a δ-function source [2]. The $M_{max}(t) \to \infty$ when $t \to t_{cr}$.

## 6. APPENDIX C

The solution of initial problem for equation (2) may be used in the parametrical form [5]

$$z = \xi + G(\xi)(x - x_0), \quad G(\xi) = g(F_0(\xi)),$$
$$F = F_0(\xi), \qquad (C1)$$

where $\xi$ is a parameter, $F_0$ is the initial value of mMF: $F_0(z) = F(z, x_0)$ for $x = x_0$, in which we have made the change of variables from "$M$" to "$z$" shown in the main text and renoted $t$ by $x$. Thus, we have from (2) $F = \beta / \sqrt{\xi - \xi_0}$ and from here:

$$\xi - \xi_0 = \beta^2 / F^2. \qquad (C2)$$

Exclude from (C1) and (C2) the parameter $\xi$, we receive the cubic equation (3), using $G(\xi)$ in the form

$$G(\xi) = C\Pi + \gamma F.$$